\documentclass[aps,prd,preprint,superscriptaddress,nofootinbib]{revtex4-2}
\usepackage{amsmath,amssymb,graphicx,bm,slashed}
\usepackage{xcolor}

\usepackage[colorlinks=true,allcolors=blue]{hyperref}
\newcommand{\Lam}{\Lambda}
\newcommand{\LUV}{\Lambda_{\rm UV}}
\newcommand{\pslash}{\slashed{p}}

\begin{document}

\title{Hermitian matrix-valued nonlocal deformation of the SU($N$) Thirring model at large $N$}

\author{J. R. Nascimento} 
\email{jroberto@fisica.ufpb.br}
\homepage{}
\affiliation{Departamento de F\'{\i}sica, Universidade Federal da Para\'{\i}ba,
Centro de Ci\^encias Exatas e da Natureza, 58051-970, Jo\~ao Pessoa, Para\'{\i}ba, Brazil}

\author{P. J. Porf\'{\i}rio}
\email{pporfirio@fisica.ufpb.br}
\homepage{https://orcid.org/0000-0002-4713-3332}
\affiliation{Departamento de F\'{\i}sica, Universidade Federal da Para\'{\i}ba,
Centro de Ci\^encias Exatas e da Natureza, 58051-970, Jo\~ao Pessoa, Para\'{\i}ba, Brazil}

\author{C. Furtado}
\email{furtado@fisica.ufpb.br}
\homepage{https://orcid.org/0000-0002-3455-4285}
\affiliation{Departamento de F\'{\i}sica, Universidade Federal da Para\'{\i}ba,
Centro de Ci\^encias Exatas e da Natureza, 58051-970, Jo\~ao Pessoa, Para\'{\i}ba, Brazil}

\date{\today}

\begin{abstract}
We investigate a Hermitian matrix-valued nonlocal deformation of the two-dimensional $SU(N)$ Thirring model at leading order in $1/N$. We impose $f(\slashed{\partial})^\ddagger=f(\slashed{\partial})$ directly on the original action; for real nonlocality scale the admissible exponential choice is $f(\slashed{\partial})=\exp[-i\slashed{\partial}/\Lambda]$. We construct and analyze the even-in-mass (paired) large-$N$ effective potential and derive its radial kernel, weak-coupling dimensional-transmutation scale, and formal Lambert-$W$ zeros. Its radial gap integral possesses a conditionally convergent infinite-cutoff limit. The pairing prescription, however, differs from the unpaired determinant of the original Hermitian action, so the reported gap and spectral zeros must be understood as properties of this explicitly defined mean-field functional. We identify the further derivation needed to obtain physical mass gaps and fermion poles from the unpaired theory.
\end{abstract}

\maketitle
\clearpage

\section{Introduction}
Nonlocal field theories have attracted renewed attention as a framework in which strictly
point-like interactions are replaced by operators carrying an intrinsic length scale,
usually through entire form factors such as $e^{\ell^2\Box}$ acting on the
fields~\cite{BasBeneito2023}. The original motivations range from Yukawa's proposal of
nonlocal fields~\cite{Yukawa} and the axiomatic construction of nonlocal scalar quantum
field theory~\cite{Efimov1967} to string field theory, where effective actions naturally
generate infinitely many derivatives~\cite{Brekke1988,Frampton1988,Eliezer1989,Ghoshal2000}.
More recently, nonlocal form factors have been employed as gauge-invariant
regulators~\cite{Evens1991}, as a route to improved ultraviolet (UV) behavior in gauge and
gravitational settings~\cite{Tomboulis1997}, and as a tool for constructing ghost-free
infinite-derivative quantum field theories~\cite{Tomboulis2015,Buoninfante2019}. The analytic
structure, causality, and unitarity constraints of such theories are discussed in
Refs.~\cite{Barnaby2008,Calcagni2008,Modesto2021}.

 In the fermionic sector, a nonlocal spin-1/2 theory was recently proposed in Ref. \cite{Nascimento2026}, in which the conventional Dirac operator leaves room for an entire form factor depending on $\slashed{\partial}$, rather than on the d'Alembertian operator that was extensively explored in the literature \cite{Krasnikov, Moffat}. Within this framework, the authors investigated the modified dispersion relation of free fermions and calculated one-loop radiative corrections in the presence of a Yukawa interaction, revealing nontrivial contributions associated with the nonlocality scale. Furthermore, they explored gauge invariance and derived a nonlocal extension of the Pauli equation. Fermions confined in the presence of a hard-wall potential have been studied within this nonlocal Dirac theory \cite{Gabriel}. A curved spacetime generalization of the nonlocal Dirac equation has been recently proposed \cite{Porfirio:2026zga}.
A distinctive feature of this formulation is the matrix-valued character of the nonlocal form factor, which requires the Dirac-matrix structure to be retained in the analysis of fermionic propagators and quantum corrections. Motivated by these developments, we extend this framework to an interacting four-fermion theory in (1+1)-dimensional spacetime \cite{Thirring1958, NJ, Klevansky}, focusing on the non-Abelian $SU(N)$ Thirring model \cite{Thirring1958} and the consequences of nonlocality for dynamical mass generation.

A nonlocal four-fermion theory based on matrix-valued
functions of the Dirac operator was recently investigated in \cite{Belchior}. In that work, a nonlocal
four-dimensional NJL/Gross--Neveu-type model  was studied
at the mean-field level, with particular attention to
dynamical mass generation and the critical coupling. Both real
and imaginary exponential form factors were considered.

The present work addresses a different physical setting:
the two-dimensional non-Abelian $SU(N)$ Thirring model.
Although the treatment of the matrix-valued Dirac
operator shares essential features with
Ref.~\cite{Belchior}, the dimensionality and the
current--current interaction lead to a distinct
large-$N$ problem. In particular, the four-fermion
coupling is dimensionless in two dimensions, allowing
the mass-gap equation to exhibit logarithmic infrared
behavior and dimensional transmutation rather than the
finite critical-coupling structure characteristic of
the four-dimensional model. We also examine the
Hermiticity condition, the contributions odd in the
auxiliary mass parameter, and the formal pole structure
of the paired nonlocal kernel.

The purpose of the present paper is to test whether this formalism survives dimensional
reduction, where the physics is qualitatively different: the four-fermion coupling is
dimensionless, the model is asymptotically free, and the dynamical mass is generated by
dimensional transmutation~\cite{GrossNeveu1974}. We consider the Thirring
model~\cite{Thirring1958} with $N$ flavors. Two points require care. First, the abelian
Thirring interaction, built from the $U(1)$ current, does not generate a mass gap: by
bosonization it only renormalizes the compactification radius of the $U(1)$
boson~\cite{Coleman1975,DashenFrishman1975}. Dynamical mass generation requires the
non-abelian $SU(N)$ currents~\cite{Witten1978}. Second, a Fierz rearrangement maps the SU($N$)
Thirring interaction exactly onto the chiral Gross--Neveu (scalar--pseudoscalar) channel, so
that the auxiliary-field machinery of Ref.~\cite{Nascimento2026} can be imported without any
ambiguity in the coupling. Because the nonlocality resides in the kinetic operator only, this
rearrangement holds identically in the nonlocal theory.

Our analysis distinguishes algebraic identities from assumptions about the
fermion determinant. We report exact kernels and controlled numerical checks
of the paired saddle, then identify which of the proposed physical conclusions
require checking the form-factor Dirac-adjoint constraint and the determinant prescription separately. In particular,
the ``poles'' discussed below are zeros of an analytically continued paired
Euclidean expression; identifying them as the spectrum of a single physical
fermion propagator is a separate dynamical question.

The paper is organized as follows. In Sec.~\ref{sec:model} we propose a nonlocal extension of the  $SU(N)$
Thirring model and perform the Fierz mapping. In Sec.~\ref{sec:gap} we derive the effective
potential and the gap equation. In Sec.~\ref{sec:kernels} we evaluate the kernels, establish
dimensional transmutation, analyze the UV behavior, and determine the pole structure. Section~\ref{sec:conclusions} contains our conclusions. Additional details on the spinor algebra and Fierz mapping, the determinant and asymptotic gap equation, and the numerical diagnostics and Lambert-$W$ roots are provided in Appendices~\ref{app:algebra}, \ref{app:gapdetails}, and \ref{app:diagnostics}, respectively.

\section{Nonlocal $SU(N)$ Thirring model and Fierz mapping}
\label{sec:model}

\subsection{Thirring model and Fierz mapping}
\label{sec:model:abelian}

 This subsection briefly reviews the main properties of the Thirring model, which was originally proposed to describe a two-dimensional current-current fermionic interaction \cite{Thirring1958}. The original (abelian) Thirring model with $N$ flavors,
\begin{equation}
\mathcal L_{U(1)}=\bar\psi_a i\slashed{\partial}\psi_a-\frac{g}{2N}\,J_\mu J^\mu,\qquad
J_\mu=\bar\psi_a\gamma_\mu\psi_a ,
\label{eq:L-U1}
\end{equation}
where $\slashed{\partial}=\gamma^{\mu}\partial_{\mu}$, $a=1,...,N$ stands for flavor index and $J_{\mu}$ is the $U(1)$ fermionic current. The Dirac gamma matrices in $(1+1)$ dimensions admit a representation in terms of the Pauli matrices, which we choose as
\begin{equation}
\gamma^0=\sigma_1,\qquad \gamma^1=i\sigma_2,\qquad \gamma_5=\gamma^0\gamma^1=-\sigma_3,
\label{eq:gammas}
\end{equation}
so that $\{\gamma^\mu,\gamma^\nu\}=2\eta^{\mu\nu}$, with metric signature $(+,-)$, and $(\gamma_5)^2=\hat{I}$. In two dimensions $[\psi]=1/2$, so that every
four-fermion coupling is dimensionless. Although Eq. \eqref{eq:L-U1} describes $N$ fermion flavors, the current–current interaction involves only the flavor-singlet $U(1)$ current. Therefore, the term Abelian Thirring model refers to the symmetry channel of the interaction rather than to the number of fermionic flavors. Upon bosonization, the interaction modifies the $U(1)$ bosonic sector, while the $SU(N)$ sector remains unaffected. Consequently, the massless theory remains conformal and does not develop a mass gap. Upon bosonization, $J_\mu$ is linear in the $U(1)$ boson and
the interaction merely rescales its kinetic term; the SU($N$) sector remains free and the
theory is conformal, without a mass gap, for every $N$~\cite{Coleman1975,DashenFrishman1975}.
The same conclusion is reached at leading order in $1/N$: a vector auxiliary field has
$\langle A_\mu\rangle=0$ by Lorentz invariance, and the scalar structures produced by a
Fierz rearrangement of Eq.~(\ref{eq:L-U1}) carry crossed flavor indices,
$(\bar\psi_a\psi_b)(\bar\psi_b\psi_a)$, whose contribution to a flavor-singlet condensate is
suppressed by $1/N$. Equation~(\ref{eq:L-U1}) is therefore not a suitable starting point for
dynamical mass generation.

We consider instead the $SU(N)$ Thirring model~\cite{DashenFrishman1975,Witten1978}
\begin{equation}
\mathcal L_{\rm Th}=\bar\psi_a i\slashed{\partial}\psi_a-\frac{g}{N}\,J^A_\mu J^{A\mu},\qquad
J^A_\mu=\bar\psi_a\gamma_\mu (T^A)_{ab}\psi_b ,
\label{eq:L-SUN}
\end{equation}
where $T^A$ are the $SU(N)$ generators in the fundamental representation, normalized as
${\rm tr}(T^AT^B)=\tfrac12\delta^{AB}$. The model is asymptotically free for $g>0$.

The distinction between the Abelian and non-Abelian interactions becomes explicit upon rearranging the spinor and flavor indices.
\label{sec:model:fierz}
For anticommuting fields and the matrices~(\ref{eq:gammas}), the two-dimensional Fierz
identity reads (its derivation and a numerical algebraic check are given in Appendix~\ref{app:fierz})
\begin{equation}
(\bar\psi_a\gamma^\mu\psi_b)(\bar\psi_b\gamma_\mu\psi_a)
=-\Big[(\bar\psi\psi)^2+(\bar\psi i\gamma_5\psi)^2\Big],
\label{eq:Fierz}
\end{equation}
where on the right-hand side all bilinears are flavor singlets,
$\bar\psi\psi\equiv\bar\psi_a\psi_a$. Combining Eq.~(\ref{eq:Fierz}) with the completeness
relation
$\sum_A (T^A)_{ab}(T^A)_{cd}=\tfrac12\big(\delta_{ad}\delta_{cb}-\tfrac1N\delta_{ab}\delta_{cd}\big)$
we obtain the exact identity
\begin{equation}
-\frac{g}{N}J^A_\mu J^{A\mu}
=\frac{g}{2N}\Big[(\bar\psi\psi)^2+(\bar\psi i\gamma_5\psi)^2\Big]
+\frac{g}{2N^2}\,J_\mu J^\mu .
\label{eq:Fierz-SUN}
\end{equation}
The first term is the chiral Gross--Neveu interaction with coupling
\begin{equation}
G=g,
\label{eq:G=g}
\end{equation}
with no dependence on the regularization: Eq.~(\ref{eq:Fierz-SUN}) is an algebraic identity
between local operators. The last term is an abelian Thirring interaction with coupling
$g/N^2$; by the argument of Sec.~\ref{sec:model:abelian} it does not contribute to mass
generation and is dropped at leading order. At the quantum level the local $SU(N)$ Thirring
model is equivalent to the chiral Gross--Neveu model up to a decoupled $U(1)$
sector~\cite{DashenFrishman1975,Witten1978}. In the nonlocal theory bosonization is not
available, and we use only the operator identity~(\ref{eq:Fierz-SUN}) together with the
leading large-$N$ approximation.

\subsection{Nonlocal deformation and Hermiticity}
\label{sec:model:nonlocal}

 Here, we propose a nonlocal extension of the Thirring model. Inspired by the seminal paper \cite{Nascimento2026}, which introduces nonlocality via entire functions of $\slashed{\partial}$. In practical terms, the nonlocal fermionic sector is obtained by promoting the Dirac operator to an entire function of
$\slashed{\partial}$, we mean, $\slashed{\partial}\rightarrow \slashed{\partial} f(\slashed{\partial})$:
\begin{equation}
\mathcal L=\bar\psi_a i\slashed{\partial} f(\slashed{\partial})\psi_a
+\frac{G}{2N}\Big[(\bar\psi\psi)^2+(\bar\psi i\gamma_5\psi)^2\Big],
\qquad f(z)=\sum_{n=0}^\infty b_n\Big(\frac{z}{\Lam}\Big)^n,\quad b_0=1,
\label{eq:L-nonlocal}
\end{equation}
with $\Lam$ the nonlocality scale. The momentum-space propagator is
\begin{equation}
S_F(p)=\frac{i}{\pslash f(\pslash)}=\frac{i\,\pslash f(\pslash)}{p^2f^2(\pslash)},
\label{eq:SF}
\end{equation}
where $f(\pslash)$ is a matrix in spinor space~\cite{Nascimento2026}.

It is useful to split $f=f_++f_-$ into its even and odd parts in $z$, see \cite{Nascimento2026} for the detailed computation. The operator
$i\slashed{\partial} f_+(\slashed{\partial})$ contains only odd powers of $\slashed{\partial}$ and anticommutes with $\gamma_5$;
it preserves both the $U(1)_A$ chiral symmetry $\psi\to e^{i\alpha\gamma_5}\psi$ and its
discrete subgroup $\psi\to\gamma_5\psi$. The operator $i\slashed{\partial} f_-(\slashed{\partial})$ contains only even
powers, $\slashed{\partial}^{2k}=\Box^k$, which commute with $\gamma_5$ and therefore act as
chirality-breaking, Wilson-like terms~\cite{Wilson}.

The even--odd decomposition also makes it necessary to distinguish chiral symmetry from Hermiticity: the latter imposes an independent restriction on the admissible form factor, as first remarked in  \cite{Porfirio:2026zga}.
\label{sec:hermiticity-selection}
We impose the admissibility condition on the kinetic form factor before introducing the auxiliary fields. The condition is applied to the original action and does not entail a separate symmetrization.
We select the exponential form factor satisfying the Dirac-adjoint Hermiticity condition,
\begin{equation}
f(\slashed{\partial})=e^{-i\slashed{\partial}/\Lam},\qquad f(\slashed{\partial})^\ddagger=f(\slashed{\partial}),
\label{eq:ff}
\end{equation}
where the Dirac adjoint of any operator $\mathcal{O}$ is defined by $\mathcal{O}^{\ddagger}\equiv\gamma^{0}\mathcal{O}^{\dagger}\gamma^0$.
For the selected exponential $f_-= -i\sin(\slashed{\partial}/\Lam)$ and $f_+=\cos(\slashed{\partial}/\Lam)$.
\begin{equation}
f_+(\slashed{\partial})=\cos\frac{\slashed{\partial}}{\Lam},\qquad f_-(\slashed{\partial})=-i\sin\frac{\slashed{\partial}}{\Lam}.
\label{eq:fpm}
\end{equation}
To derive this restriction explicitly, we use the Dirac-adjoint operation on differential operators, with integration by parts included.
The action (\ref{eq:L-nonlocal}) need not be replaced by a manifestly symmetrized kinetic term. Instead, one imposes self-adjointness directly on the matrix-valued form factor with respect to the Dirac adjoint, including the integration-by-parts operation. With boundary terms assumed to vanish, the Dirac pairing obeys
\begin{equation}
\langle\chi,\mathcal O\psi\rangle_D
=\langle\mathcal O^\ddagger\chi,\psi\rangle_D,\qquad
\langle\chi,\psi\rangle_D=\int d^2x\,\bar\chi\psi .
\end{equation}
In particular, $\slashed{\partial}^\ddagger=-\slashed{\partial}$ whereas $(i\slashed{\partial})^\ddagger=i\slashed{\partial}$. Because an entire function of $\slashed{\partial}$ commutes with $i\slashed{\partial}$, the kinetic operator satisfies
\begin{equation}
[i\slashed{\partial} f(\slashed{\partial})]^\ddagger
=f(\slashed{\partial})^\ddagger i\slashed{\partial};
\qquad
f(\slashed{\partial})^\ddagger=f(\slashed{\partial})
\Longrightarrow [i\slashed{\partial} f(\slashed{\partial})]^\ddagger=i\slashed{\partial} f(\slashed{\partial}).
\label{eq:hermiticity}
\end{equation}

For a power series $f(\slashed{\partial})=\sum_{n\geq0}b_n(\slashed{\partial}/\Lam)^n$, the constraint implies $b_n=(-1)^n b_n^*$ since $\Lam$ is real. Thus even coefficients are real and odd coefficients purely imaginary. The selected form factor obeys
\begin{equation}
\left(e^{-i\slashed{\partial}/\Lam}\right)^\ddagger=e^{-i\slashed{\partial}/\Lam}.
\label{eq:exponential-adjoint}
\end{equation}

 The real exponential $e^{-\slashed{\partial}/\Lam}$ instead satisfies $(e^{-\slashed{\partial}/\Lam})^\ddagger=e^{+\slashed{\partial}/\Lam}\neq e^{-\slashed{\partial}/\Lam}$ and is therefore excluded from the physical model. This real form factor will not be considered in our following investigations. These are formal adjoint statements subject to suitable common operator domains and vanishing boundary contributions. Although the selected form factor satisfies the Hermiticity constraint and preserves parity, these properties do not necessarily imply chiral invariance. Indeed, expanding the nonlocal kinetic operator in powers of the inverse nonlocality scale yields
\begin{equation}
i\slashed{\partial}e^{-i\slashed{\partial}/\Lambda}
=i\slashed{\partial}
+\frac{\Box}{\Lambda}
-\frac{i\slashed{\partial}\Box}{2\Lambda^2}
+\mathcal{O}(\Lambda^{-3}).
\end{equation}
While the terms containing odd powers of ($\slashed{\partial}$) anticommute with ($\gamma_5$), those involving even powers commute with it and therefore explicitly break chiral symmetry. In particular, the contribution ($\Box/\Lambda$) represents the leading chirality-breaking correction. Consequently, the nonlocal theory is Hermitian but not chirally invariant at finite ($\Lambda$), even in the absence of a bare fermion mass. Chiral symmetry is recovered in the local limit ($\Lambda\rightarrow\infty$).

\subsection{Auxiliary fields and large-$N$ effective action}
\label{sec:model:HS}
To investigate dynamical mass generation in the nonlocal $SU(N)$ Thirring model, we employ the Hubbard–Stratonovich transformation \cite{Stra, Hubbard}, which allows the four-fermion interaction to be expressed in terms of auxiliary bosonic fields. Specifically, we introduce a scalar field $\sigma$ and a pseudoscalar field $\pi$, associated with the fermionic bilinears $\bar\psi\psi$ and $\bar\psi i\gamma_5\psi$, respectively. This transformation renders the fermionic action quadratic, allowing the fermionic degrees of freedom to be integrated out and the resulting effective action to be analyzed systematically in the large-$N$ limit. The homogeneous scalar background then provides a mass parameter whose value is determined self-consistently through the stationary condition of the effective potential. The resulting Hubbard–Stratonovich Lagrangian takes the form
\begin{equation}
\mathcal L_{\rm HS}=\bar\psi_a\big(i\slashed{\partial} f(\slashed{\partial})-\sigma-i\gamma_5\pi\big)\psi_a
-\frac{N}{2G}\big(\sigma^2+\pi^2\big).
\label{eq:LHS}
\end{equation}
 Varying the action with respect to the auxiliary fields $\sigma$ and $\pi$, respectively, we obtain a couple of algebraic field equations
\begin{equation}
\sigma=-\frac{G}{N}\bar{\psi}\psi,
\qquad
\pi=-\frac{G}{N}\bar{\psi}i\gamma_5\psi.
\end{equation}
Now, putting them back into Eq.\eqref{eq:LHS}, one recovers Eq.~(\ref{eq:L-nonlocal}); the expectation
values parametrize $\langle\bar\psi\psi\rangle$ and $\langle\bar\psi i\gamma_5\psi\rangle$. In
the background $(\sigma,\pi)$ the Dirac operator is
$D_{\sigma\pi}=i\slashed{\partial} f(\slashed{\partial})-\sigma-i\gamma_5\pi$, and integrating out the fermions gives
\begin{equation}
S_{\rm eff}[\sigma,\pi]=-N\!\int\! d^2x\,\frac{\sigma^2+\pi^2}{2G}-iN\,{\rm Tr}\ln D_{\sigma\pi},
\label{eq:Seff}
\end{equation}
where Tr runs over spacetime and Dirac indices. Fluctuations of $\sigma,\pi$ are suppressed by
$1/N$. Since the nonlocal operator preserves parity and $\pi$ is parity odd, the effective
potential is even in $\pi$ and $\bar\pi=0$ is always a stationary point. We therefore take
the homogeneous saddle $\bar\sigma=M$, $\bar\pi=0$, for which the gap equation coincides with
that of the (discrete) Gross--Neveu model at leading order.

Two remarks on symmetry are in order. At finite $N$ the continuous U(1)$_A$ symmetry cannot be
broken spontaneously in two dimensions~\cite{Coleman1973}; at large $N$ the fermion
nevertheless acquires the mass $M$ while $\langle\bar\psi\psi\rangle$ vanishes, in the sense
discussed by Witten~\cite{Witten1978}. The homogeneous auxiliary background $M$ is a mean-field mass parameter; its identification with a physical fermion mass requires the pole analysis of the unpaired propagator. Moreover, for $f_-\neq0$ the chiral
symmetry is broken explicitly at the scale $\Lam$, so that ``chiral limit'' refers below to
the tuned limit defined in Sec.~\ref{sec:gap}.

\section{Effective potential and matrix-valued gap equation}
\label{sec:gap}

\subsection{Euclidean effective potential and Dirac trace}

Following the Euclidean prescription discussed in details in Ref. \cite{Nascimento2026}, we formulate the quantum corrections directly in Euclidean space rather than performing a conventional Wick rotation of the Minkowski loop integrals. Indeed, in nonlocal field theories, entire form factors may exhibit exponential growth along certain directions of the complex momentum plane, preventing the vanishing of contour contributions at infinity. Consequently, the equivalence between Wick rotation and Euclidean integration is not generally guaranteed. We therefore adopt a Euclidean formulation of the fermionic effective action, with the nonlocal kinetic operator defined directly in Euclidean signature.

At the homogeneous saddle $\bar{\sigma}=M, \bar{\pi}=0$, the Euclidean effective action is $\Gamma_E=NV_2V_{\mathrm{eff}}(M)$, where $V_2$ denotes the Euclidean two-volume.
Because the form factor is matrix valued and not even in $\pslash$, the unsymmetrized
Euclidean determinant contains a part that is odd in $M$. For the eigenvalues $\pm p$ of $\pslash$ one finds (Appendix~\ref{app:gapdetails})
\begin{equation}
\det_D\big[M+i\pslash f(\pslash)\big]=M^2+p^2+2Mp\sin(p/\Lam).
\label{eq:detodd}
\end{equation}

For clarity, the spinor-space determinant follows without replacing the form factor by a scalar. Diagonalizing $\pslash$ with eigenvalues $+p$ and $-p$, and writing $\theta=p/\Lam$, one obtains
\begin{align}
\det_D[M+i\pslash e^{-i\pslash/\Lam}]
&=(M+ip e^{-i\theta})(M-ip e^{i\theta})\nonumber\\
&=M^2+p^2+2Mp\sin\theta,\label{eq:det-detail}\\
\det_D[M-i\pslash e^{-i\pslash/\Lam}]
&=M^2+p^2-2Mp\sin\theta.\label{eq:det-partner}
\end{align}
Thus pairing amounts to replacing the logarithm of the first factor by the logarithm of the geometric mean of these two different determinants.

The odd part is the Euclidean image of the chirality-breaking operator $i\slashed{\partial} f_-$. The treatment of matrix-valued fermionic form factors
in a four-fermion mean-field effective action was
previously developed in Ref.~\cite{Belchior}.
We employ the corresponding paired-determinant
construction in the present two-dimensional setting.
However, because the selected form factor is not
even in $\slashed p$, we explicitly distinguish the
paired contribution from the unpaired fermionic
determinant and isolate their difference below. To start with, we define the mean-field potential from the paired
determinant,
\begin{equation}
\ln\det D_E(p)\ \longrightarrow\ \frac12\ln{\det}_D\big[D_+(p)D_-(p)\big],\qquad
D_\pm(p)=M\pm i\pslash f(\pslash),
\label{eq:paired}
\end{equation}
which retains exactly the $M$-even part. This replacement defines a distinct
mean-field functional and is not, by itself, an identity for the original
fermionic path integral. In particular, taking the geometric average of
determinants is not equivalent to adding a conventional local mass counterterm.
 The appearance of a term odd in the auxiliary field $M$ is a direct consequence of the explicit chiral-symmetry breaking induced by the nonlocal kinetic operator. Although the selected form factor satisfies the Hermiticity constraint, this condition does not eliminate the mass-odd contribution to the fermionic determinant. Consequently, canceling its leading contribution through an additive mass counterterm does not restore chiral symmetry. The paired-determinant prescription therefore defines a distinct mean-field approximation whose predictions must be distinguished from those obtained from the full fermionic determinant. For the admitted oscillatory form factor the
discarded part is real; its leading effect is an additive mass renormalization
$m_0=(G\Lam/\pi)\big[1-\cos(\LUV/\Lam)\big]$, which we cancel with a bare-mass counterterm,
exactly as the additive mass renormalization of Wilson fermions is tuned to reach the chiral
point~\cite{Wilson}. This cancels only the leading tadpole, not the higher odd powers of $M$.
Writing $A_0=M^2+p^2$ and $b=2Mp\sin(p/\Lam)$, the difference between
 the original unpaired and paired fermionic contributions is
\begin{equation}
V_{\rm odd}(M)=-\frac12\int\frac{d^2p}{(2\pi)^2}
\ln\!\left(\frac{A_0+b}{A_0-b}\right).
\label{eq:oddexp}
\end{equation}
This formula is obtained by subtracting the even-in-$M$ average of the two
logarithmic Dirac determinants from the unpaired logarithmic determinant.
Its expansion, the cutoff-dependent linear term, and the limits of additive
mass tuning are derived in Appendix~\ref{app:det}.  In particular, canceling
the first derivative at $M=0$ does not remove the remaining mass-odd terms.
Since $f(\pslash)$ commutes with $\pslash$,
\begin{equation}
D_+(p)D_-(p)=M^2+p^2f^2(\pslash),
\end{equation}
and the effective potential per flavor is
\begin{equation}
V_{\rm eff}(M)=\frac{M^2}{2G}-\frac12\int\!\frac{d^2p}{(2\pi)^2}\,
\ln{\det}_D\big[p^2f^2(\pslash)+M^2\big].
\label{eq:Veff}
\end{equation}
The nonlocality does not enter as a multiplicative correction to the local determinant; it
changes the matrix structure inside the logarithm through $f^2(\pslash)$, and the scalar
function that controls the potential can be identified only after the Dirac determinant is
evaluated.

To extract the scalar gap kernel from the matrix-valued determinant in Eq.~(\ref{eq:Veff}), we now invert its Dirac-space operator without treating the form factor as a scalar. With the Euclidean Clifford algebra $\{\gamma^\mu,\gamma^\nu\}=2\delta^{\mu\nu}$,
$\pslash^2=p^2\hat{I}$. We decompose
\begin{equation}
f^2(\pslash)=a_1(p)\hat{I}+a_2(p)\pslash,\qquad p=\sqrt{p^2},
\label{eq:a1a2}
\end{equation}
 where the nonlocal coefficients in the momenta space are explicitly given by $a_1(p)=\cos\left(\frac{2p}{\Lambda}\right)$ and $a_2(p)=-\dfrac{i}{p}\sin\left(\frac{2p}{\Lambda}\right)$ for the complex exponential form factor,
and write
\begin{equation}
\mathcal D(p)=p^2f^2(\pslash)+M^2=A(p)\hat{I}+B(p)\pslash,\qquad
A=p^2a_1+M^2,\quad B=p^2a_2 .
\label{eq:AB}
\end{equation}
Using $\pslash^2=p^2\hat{I}$,
\begin{equation}
\mathcal D^{-1}(p)=\frac{A(p)\hat{I}-B(p)\pslash}{A^2(p)-B^2(p)p^2},\qquad
{\rm tr}_D\,\mathcal D^{-1}(p)=\frac{2A(p)}{A^2(p)-B^2(p)p^2},
\label{eq:Dinv}
\end{equation}
since only the identity contributes to the trace and ${\rm tr}_D\hat{I}=2$.
Equivalently, the spectral projectors $P_\pm=(\hat I\pm\pslash/p)/2$ yield
\begin{equation}
\mathcal D^{-1}=\frac{P_+}{M^2+p^2 e^{-2ip/\Lam}}
+\frac{P_-}{M^2+p^2 e^{+2ip/\Lam}},\qquad
\frac12\mathrm{tr}_D\,\mathcal D^{-1}
=\mathrm{Re}\frac{1}{M^2+p^2e^{-2ip/\Lam}}.
\label{eq:projector-inverse}
\end{equation}

\subsection{Gap equation and local limit}

Stationarity, $\partial V_{\rm eff}/\partial M=0$, gives for the nontrivial solution
\begin{equation}
\frac1G=\int\!\frac{d^2p}{(2\pi)^2}\,\frac{2A(p)}{A^2(p)-B^2(p)p^2}
=\frac1\pi\int_0^{\LUV}\!dp\,p\,K_f(p;M,\Lam),
\label{eq:gap}
\end{equation}
where an O(2)-invariant cutoff $\LUV$ has been introduced and
\begin{equation}
K_f(p;M,\Lam)=\frac{p^2a_1(p)+M^2}{\big[p^2a_1(p)+M^2\big]^2-p^6a_2^2(p)} .
\label{eq:kernel}
\end{equation}
Equation~(\ref{eq:gap}) is the two-dimensional analogue of Eq.~(75) of
Ref.~\cite{Nascimento2026}. The coupling $G$ is dimensionless.

The normalization and infrared structure of the gap equation can be checked directly against the local theory. For $f\to1$, $K_{\rm loc}=1/(p^2+M^2)$ and
\begin{equation}
\frac1G=\frac1{2\pi}\ln\Big(1+\frac{\LUV^2}{M^2}\Big),
\qquad M=\frac{\LUV}{\sqrt{e^{2\pi/G}-1}}\simeq\LUV\,e^{-\pi/G},
\label{eq:local}
\end{equation}
the cutoff-regularized Gross--Neveu gap equation~\cite{GrossNeveu1974,MosheZinnJustin2003}.
A nontrivial solution exists for every $G>0$: this is dimensional transmutation in the
asymptotically free theory.

\section{Kernels, dimensional transmutation, and pole structure}
\label{sec:kernels}

\subsection{Nonlocal kernel and infrared expansion}
\label{sec:kernels:explicit}

Having established the matrix-valued structure of the effective potential and the corresponding gap equation, we now determine the explicit momentum dependence of the nonlocal kernel. For the Hermitian form factor \(f(\slashed p)=e^{-i\slashed p/\Lambda}\), the identity \(\slashed p^2=p^2\hat I\) allows us to express its square as
\[
f^2(\slashed p)=e^{-2i\slashed p/\Lambda}
=\cos\left(\frac{2p}{\Lambda}\right)\hat I
-i\frac{\slashed p}{p}\sin\left(\frac{2p}{\Lambda}\right).
\]
Substituting this expression into the general kernel equation \eqref{eq:kernel}, we obtain
\begin{equation}
K_f(p;M,\Lam)=\frac{p^2\cos(2p/\Lam)+M^2}{p^4+2M^2p^2\cos(2p/\Lam)+M^4}.
\label{eq:KII}
\end{equation}

The kernel can also be written as $K_f=\operatorname{Re}[M^2+p^2e^{-2ip/\Lam}]^{-1}$ and is regular for real $p$ away from parameter choices at which its denominator vanishes. It need not be pointwise positive.
\begin{figure}[t]
\centering\includegraphics[width=0.48\textwidth]{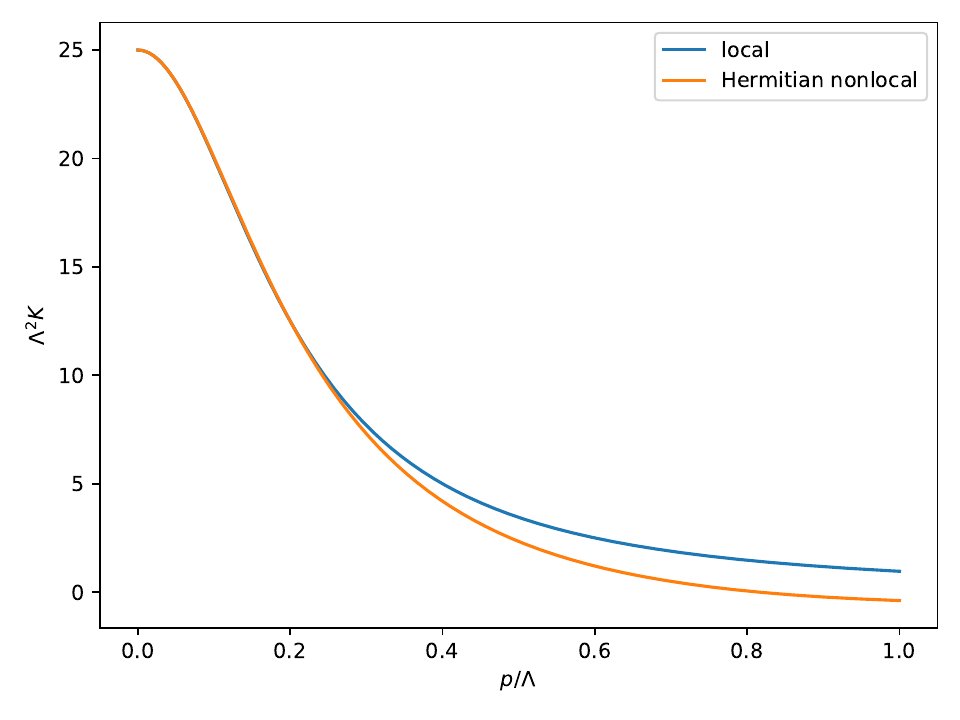}
\caption{Paired gap kernels for the local theory and the admitted Hermitian nonlocal form factor, at $M/\Lam=0.2$. The Hermitian nonlocal kernel is not pointwise positive and can become negative at finite momentum; the plotted quantity is the paired gap-equation kernel, not a probability density.}
\label{fig:kernels}
\end{figure}
The approach to the local theory is seen by expanding this exact kernel at fixed momentum and mass. For fixed $p$ and $M$, the large-$\Lam$ expansion gives
\begin{equation}
K_f(p;M,\Lam)=\frac1{p^2+M^2}-\frac{2p^4(p^2-M^2)}{\Lam^2(p^2+M^2)^3}+O(\Lam^{-4}).
\label{eq:IR}
\end{equation}

The leading correction suppresses the integrand when $p>M$.

\subsection{Dimensional transmutation and ultraviolet behavior}
\label{sec:kernels:transmutation}

At $M=0$ the kernel reduces to $K_f=\cos(2p/\Lam)/p^2$, so the
right-hand side of Eq.~(\ref{eq:gap}) diverges logarithmically at $p\to0$, exactly as in
the local theory. The logarithmic small-mass behavior implies a nonzero weak-coupling solution
of the paired saddle for arbitrarily small $G>0$ (with a fixed finite cutoff). (A finite ``critical coupling'' can only be
produced by an infrared regulator $m_{\rm IR}$, and then depends on it logarithmically,
$G_c\simeq\pi/\ln(\LUV/m_{\rm IR})$; see Table~\ref{tab:appIR} in Appendix~\ref{app:diagnostics}.)

The nonlocal deformation instead shifts the scale generated by transmutation. Writing
$K_f=K_{\rm loc}+(K_f-K_{\rm loc})$ in Eq.~(\ref{eq:gap}), the second piece is infrared finite
as $M\to0$, and for $M\ll\Lam,\LUV$ we obtain
\begin{equation}
\frac\pi G=\ln\frac{\LUV}{M}+c_f(s)+O\!\Big(\frac{M^2}{\Lam^2}\ln\frac{\Lam}{M}\Big),\qquad s\equiv\frac{\LUV}{\Lam},
\label{eq:transmutation}
\end{equation}
The logarithm can be isolated explicitly before taking the small-mass limit:
\begin{align}
\frac\pi G&=\int_0^{\LUV}\!dp\,\frac{p}{p^2+M^2}
+\int_0^{\LUV}\!dp\,p\,[K_f(p;M,\Lam)-K_{\rm loc}(p;M)]\nonumber\\
&=\frac12\ln\left(1+\frac{\LUV^2}{M^2}\right)
+\int_0^{2s}\!dt\,\frac{\cos t-1}{t}+o(1),\quad M\to0.
\label{eq:transmutation-steps}
\end{align}
The finite part of the nonlocal kernel is
\begin{equation}
c_f(s)=-\int_0^{2s}\frac{1-\cos t}{t}\,dt=-\operatorname{Cin}(2s).
\label{eq:cf}
\end{equation}
  
\begin{equation}
M_f=\LUV e^{c_f(s)}e^{-\pi/G}.
\label{eq:Mf}
\end{equation}
At $s=1$, $e^{c_f}=0.4285$. Equation~(\ref{eq:Mf}) agrees with the small-mass numerical solutions as documented in Table~\ref{tab:M}.

This result provides the principal distinction from the four-dimensional
matrix-valued nonlocal four-fermion model of Ref.~\cite{Belchior}.
In the latter setting, the mean-field gap equation displays a finite
critical-coupling threshold, whereas the two-dimensional radial measure
produces a logarithmic infrared contribution in the paired $SU(N)$ saddle.
For $M\ll\Lam,\LUV$, Eq.~\eqref{eq:Mf} makes the difference explicit:
$M_f/\LUV\propto\exp(-\pi/G)$, with the nonlocality-dependent prefactor
$\exp[-\operatorname{Cin}(2\LUV/\Lam)]$.
This comparison concerns the specified paired functional; it does not
establish the corresponding mass-generation law for the unpaired
fermionic determinant.

\begin{figure}[t]
\centering\includegraphics[width=0.48\textwidth]{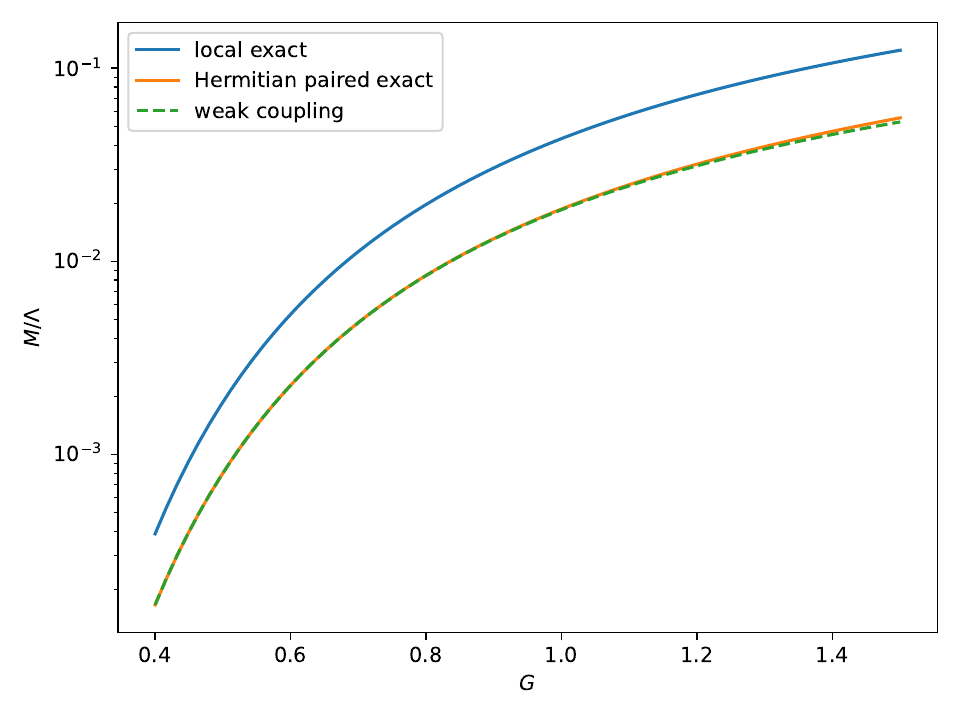}
\caption{Paired dynamical mass as a function of $G$ for the local theory and the Hermitian nonlocal model at $\LUV=\Lam$. The nonlocal curve represents the mass parameter determined by the even-in-mass paired mean-field gap equation, not an independently established physical fermion mass gap of the original unpaired Hermitian theory. The dashed curve shows its weak-coupling approximation.}
\label{fig:MG}
\end{figure}
\begin{table}[t]
\caption{Numerical solutions of the paired gap equation and weak-coupling approximation for the admitted form factor; $\LUV=\Lam$.}
\label{tab:M}
\begin{ruledtabular}
\begin{tabular}{ccc}
$G$ & paired gap (exact) & weak-coupling asymptotic \\
0.4 & $1.664\times10^{-4}$ & $1.664\times10^{-4}$\\
0.6 & $2.281\times10^{-3}$ & $2.280\times10^{-3}$\\
0.8 & $8.462\times10^{-3}$ & $8.443\times10^{-3}$\\
1.0 & $1.868\times10^{-2}$ & $1.852\times10^{-2}$
\end{tabular}
\end{ruledtabular}
\end{table}

The same kernel also determines what happens when the radial ultraviolet cutoff is removed. Unlike the local logarithmic divergence, its large-momentum behavior is oscillatory.
\label{sec:kernels:UV}
For $p\gg\Lam$ and fixed $M$, $K_f\simeq\cos(2p/\Lam)/p^2$, which gives a conditionally convergent radial integral under the stated radial-cutoff prescription. Since $\operatorname{Cin}(x)=\gamma_E+\ln x-\operatorname{Ci}(x)$, the weak-coupling mass becomes
\begin{equation}
M_f\simeq\frac{e^{-\gamma_E}}2\Lam e^{-\pi/G}\qquad(\LUV\to\infty).
\label{eq:MII-inf}
\end{equation}
This does not prove UV finiteness of the entire theory, nor independence of the regularization and integration order.
\begin{figure}[t]
\centering\includegraphics[width=0.48\textwidth]{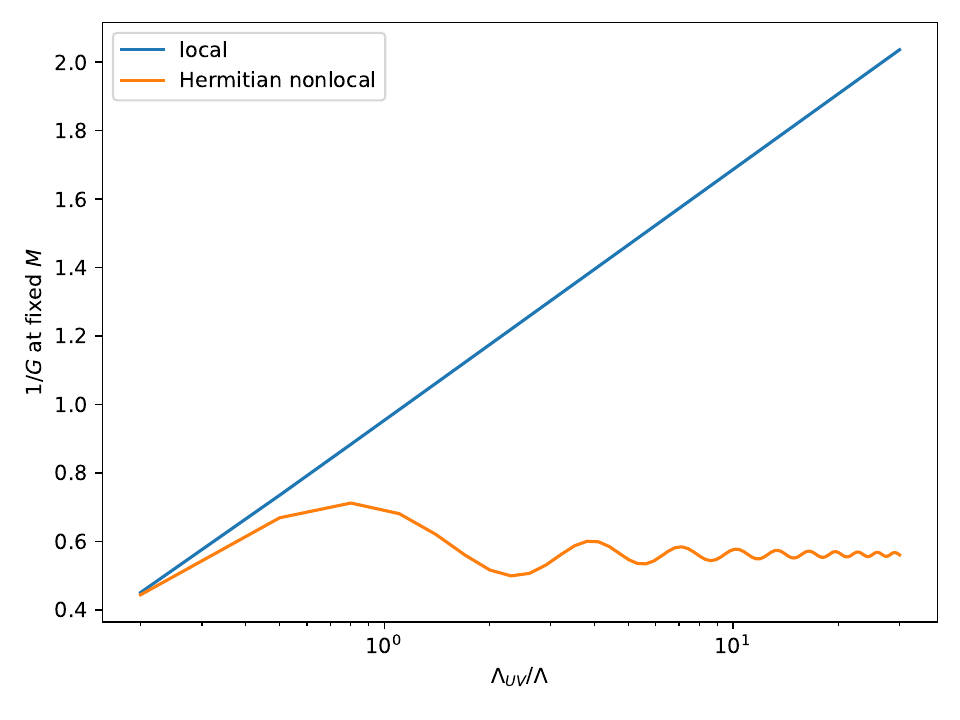}
\caption{Cutoff dependence of the local and admitted oscillatory paired gap integrals at fixed $M/\Lam=0.05$. The oscillatory nonlocal integral has a conditionally convergent limit for the specified sharp radial-cutoff prescription; the plot does not establish regulator-independent ultraviolet finiteness of the theory.}
\label{fig:cutoff}
\end{figure}

\subsection{Quasiparticle poles}
\label{sec:kernels:poles}

We follow the analytic-continuation strategy for matrix-valued Dirac form factors discussed in Ref.~\cite{Nascimento2026}, rather than introducing an independent prescription. At zero spatial momentum the Euclidean eigenvalues $\pm p$ continue to $\pm iq$, so that $\cos(2p/\Lam)\to\cosh(2q/\Lam)$. With $p^2\to-q^2$, the denominator of Eq.~(\ref{eq:KII}) becomes $q^4-2M^2q^2\cosh(2q/\Lam)+M^4$, giving the factorization
\begin{equation}
\mathcal D_f(q)=\big(q^2-M^2 e^{2q/\Lam}\big)\big(q^2-M^2 e^{-2q/\Lam}\big).
\label{eq:DII}
\end{equation}

The positive-frequency branch equations have the Lambert-function solutions, whose derivation and formal residue analysis are detailed in Appendix~\ref{app:diagnostics} (specifically, Sec.~\ref{app:poles}):
\begin{align}
q_-&=\Lam W_0(M/\Lam)\simeq M-M^2/\Lam,\nonumber\\
q_+&=-\Lam W_0(-M/\Lam)\simeq M+M^2/\Lam,\nonumber\\
q_h&=-\Lam W_{-1}(-M/\Lam).\label{eq:polesII}
\end{align}
In particular, the increasing branch solves $q=M e^{q/\Lam}$. On setting $y=-q/\Lam$ this becomes $y e^y=-M/\Lam$, so $q=-\Lam W_k(-M/\Lam)$. The two real branches $k=0,-1$ exist only for $0<M/\Lam\leq 1/e$. Differentiating $h(q)=q-M e^{q/\Lam}$ at a zero gives $h'(q)=1-q/\Lam$, establishing the relative sign change of the formal scalar residues.

For $M=0.1\Lam$, $(q_-,q_+,q_h)=(0.0913,0.1118,3.577)\Lam$. The two latter roots exist on the real axis for $0<M<\Lam/e$, coalescing at $M=\Lam/e$. The heavy-branch scalar denominator has derivative proportional to $1-q/\Lam$, which changes sign between $q_+$ and $q_h$. These are zeros and relative residues of the analytically continued paired kernel, not by themselves physical particle or ghost poles of the unpaired Hermitian theory.
\begin{figure}[t]
\centering\includegraphics[width=0.48\textwidth]{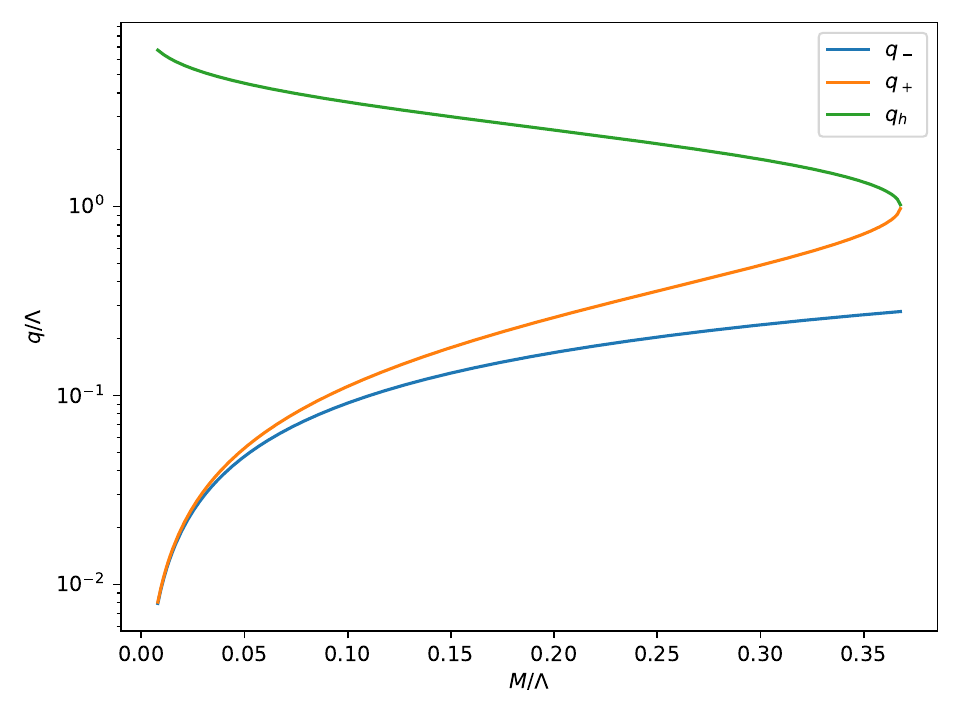}
\caption{Lambert-$W$ zeros of the admitted nonlocal paired kernel. The heavy branch is formally of opposite scalar residue sign. The $q_+$ and $q_h$ branches coalesce at $M/\Lam=1/e$. These zeros and relative scalar residues arise from the analytically continued paired kernel and do not, by themselves, establish physical fermionic poles or ghosts in the original unpaired Hermitian theory.}
\label{fig:poles}
\end{figure}

\section{Conclusions}\label{sec:conclusions}

We have investigated a Hermitian matrix-valued nonlocal deformation of the
 two-dimensional $SU(N)$ Thirring model at leading order in the large-$N$
 expansion. The Fierz rearrangement identifies scalar and pseudoscalar
 auxiliary-field channels, while imposing Dirac-adjoint Hermiticity selects
 an oscillatory exponential form factor within the ansatz considered. The
 resulting kinetic operator explicitly breaks axial chiral symmetry.

The homogeneous fermionic determinant contains contributions odd in the
 auxiliary mass. We have separated these contributions from the
 even-in-mass, paired effective potential used for the gap analysis.
 Within that precisely specified paired prescription, the two-dimensional
 gap equation exhibits logarithmic infrared behavior and dimensional
 transmutation, with a nonlocal modification of the generated mass scale.
 Numerical solutions are consistent with the weak-coupling asymptotics.
 The radial kernel also has a conditionally convergent infinite-cutoff
 limit, although this does not establish ultraviolet finiteness of the
 full quantum theory. The formal Lambert-$W$ roots similarly require a
 controlled continuation and a spectral analysis before they can be
 assigned a physical interpretation.

Compared with the four-dimensional nonlocal four-fermion theory of
 Ref.~\cite{Belchior}, the distinctive result here is the weak-coupling
 dimensional-transmutation structure of the two-dimensional paired
 saddle, rather than the determinant technique itself. Two natural
 continuations are to determine how the full, unpaired determinant
 modifies the vacuum and the mass gap, and to compute the scalar and
 pseudoscalar collective-mode propagators. The latter would reveal how
 the explicitly broken axial symmetry affects the pseudoscalar spectrum.

\begin{acknowledgments}
The authors thank the Coordena\c{c}\~ao de Aperfei\c{c}oamento de Pessoal de N\'{\i}vel
Superior (CAPES), the Para\'{\i}ba State Research Foundation (FAPESQ-PB), and the Conselho
Nacional de Desenvolvimento Cient\'{\i}fico e Tecnol\'ogico (CNPq) for financial support.
P.~J.~P. thanks CNPq (grant No.~307011/2026-7).
C.~F. thanks CNPq (grant No.~311781/2021-7).
\end{acknowledgments}

\appendix
\begingroup
\section{Spinor algebra and SU($N$) Fierz mapping}\label{app:algebra}
\subsection{Conventions and spinor basis}\label{app:conv}
We use $\eta_{\mu\nu}=\operatorname{diag}(1,-1)$,
$\gamma^0=\sigma_1$, $\gamma^1=i\sigma_2$, and
$\gamma_5=\gamma^0\gamma^1=-\sigma_3$, whence
$(\gamma_5)^2=\hat I$ and $\{\gamma_5,\gamma^\mu\}=0$.
The basis $\Gamma_A=\{\hat I,\gamma^0,\gamma^1,i\gamma_5\}$
is used with its trace-dual basis $\Gamma^A$,
$\operatorname{tr}(\Gamma_A\Gamma^B)=2\delta_A{}^B$.
The Euclidean gamma matrices are Hermitian and
$\slashed p^2=p^2\hat I$, so that $\slashed p$ has eigenvalues $\pm p$.
Flavor generators satisfy $\operatorname{tr}(T^AT^B)=\delta^{AB}/2$.

\subsection{Fierz identities and SU($N$) decomposition}\label{app:fierz}
Spinor completeness gives
\begin{equation}\label{eq:app-complete}
\delta_{\alpha\beta}\delta_{\gamma\delta}
=\frac12\sum_A(\Gamma_A)_{\alpha\delta}
 (\Gamma^A)_{\gamma\beta}.
\end{equation}
For Grassmann fields, reordering the bilinears provides the minus sign in
\begin{equation}\label{eq:app-fierz-rearrange}
(\bar\psi_a X\psi_b)(\bar\psi_b Y\psi_a)
=-X_{\alpha\beta}Y_{\gamma\delta}
 (\bar\psi_{a\alpha}\psi_{a\delta})
 (\bar\psi_{b\gamma}\psi_{b\beta}).
\end{equation}
Expanding $X\otimes Y$ in the complete Dirac basis leads to
\begin{align}
(\bar\psi_a\gamma^\mu\psi_b)(\bar\psi_b\gamma_\mu\psi_a)
&=-\big[(\bar\psi\psi)^2+(\bar\psi i\gamma_5\psi)^2\big],\label{eq:app-fierz1}\\
(\bar\psi_a\psi_b)(\bar\psi_b\psi_a)
+ (\bar\psi_a i\gamma_5\psi_b)(\bar\psi_b i\gamma_5\psi_a)
&=-(\bar\psi\gamma^\mu\psi)(\bar\psi\gamma_\mu\psi),\label{eq:app-fierz2}\\
(\bar\psi_a\psi_b)(\bar\psi_b\psi_a)
- (\bar\psi_a i\gamma_5\psi_b)(\bar\psi_b i\gamma_5\psi_a)
&=-\big[(\bar\psi\psi)^2+(\bar\psi i\gamma_5\psi)^2\big].\label{eq:app-fierz3}
\end{align}
Here the unsuffixed bilinears denote flavor singlets. The tensor coefficients
can also be checked by solving the
$16\times16$ Dirac-basis linear system. This provides an independent algebraic check of the Fierz identities used in the main text.
With
$\sum_AT^A_{ab}T^A_{cd}=(\delta_{ad}\delta_{cb}-\delta_{ab}\delta_{cd}/N)/2$,
\begin{equation}\label{eq:app-current}
J_\mu^AJ^{A\mu}=-\frac12\big[(\bar\psi\psi)^2
 +(\bar\psi i\gamma_5\psi)^2\big]-\frac1{2N}J_\mu J^\mu.
\end{equation}
This is the source of the leading large-$N$ identification $G=g$;
the nonlocal kinetic operator does not enter the spinor rearrangement.

\section{Determinant and asymptotic gap equation}\label{app:gapdetails}
\subsection{Unpaired determinant and mass-odd terms}\label{app:det}

We derive the difference between the unsymmetrized fermionic determinant
obtained from the homogeneous Hubbard--Stratonovich saddle and the paired
mean-field functional introduced in Eq.~\eqref{eq:paired}.  Throughout this
subsection, $M$ is a constant scalar auxiliary background and the
pseudoscalar background vanishes.  In the adopted Euclidean convention,
\begin{equation}
D_\pm(p)=M\hat I\pm i\pslash e^{-i\pslash/\Lam},
\qquad \pslash^2=p^2\hat I, \qquad p=\sqrt{p_\mu p_\mu}.
\end{equation}
The spinor spectral projectors
$P_\pm=\tfrac12(\hat I\pm\pslash/p)$, defined at $p\neq0$,
diagonalize both operators.  Since the eigenvalues of $\pslash$ are
$+p$ and $-p$, the two determinants are
\begin{align}
\det_D D_+(p)
 &=\left(M+ip e^{-ip/\Lam}\right)
   \left(M-ip e^{ip/\Lam}\right)\nonumber\\
 &=M^2+p^2+2Mp\sin(p/\Lam),\label{eq:app-unpaired}\\
\det_D D_-(p)
 &=\left(M-ip e^{-ip/\Lam}\right)
   \left(M+ip e^{ip/\Lam}\right)\nonumber\\
 &=M^2+p^2-2Mp\sin(p/\Lam).\label{eq:app-partner}
\end{align}
Thus the determinants are generally different for finite $\Lam$.  On the
other hand, finite-dimensional multiplicativity in spinor space gives
\begin{align}
\det_D[D_+(p)D_-(p)]
 &=\det_D D_+(p)\det_D D_-(p)\nonumber\\
 &=\big[M^2+p^2+2Mp\sin(p/\Lam)\big]
   \big[M^2+p^2-2Mp\sin(p/\Lam)\big]\nonumber\\
 &=M^4+p^4+2M^2p^2\cos(2p/\Lam).
\label{eq:app-paired}
\end{align}
The full homogeneous fermionic contribution to the Euclidean effective
potential is 
\begin{equation}
V_{\rm full}=\int \frac{d^2p}{(2\pi)^2}\,\ln\det_D D_+,
\end{equation}
whereas the paired
contribution is 
\begin{equation}
V_{\rm paired}=-\tfrac12\int \frac{d^2p}{(2\pi)^2}\,\ln\det_D(D_+D_-).
\end{equation}
Their
difference, denoted by $V_{\rm odd}=V_{\rm full}-V_{\rm paired}$, is
\begin{align}
V_{\rm odd}(M)
 &=-\frac12\int\frac{d^2p}{(2\pi)^2}
 \ln\!\left(\frac{M^2+p^2+2Mp\sin(p/\Lam)}
 {M^2+p^2-2Mp\sin(p/\Lam)}\right)\nonumber\\
 &=-\int\frac{d^2p}{(2\pi)^2}
 \left[\frac{b}{A_0}+\frac{b^3}{3A_0^3}
 +\frac{b^5}{5A_0^5}+\cdots\right],
 \label{eq:app-odd}
\end{align}
where $A_0=M^2+p^2$ and $b=2Mp\sin(p/\Lam)$.  The second line follows
from $\ln[(1+x)/(1-x)]=2\sum_{n\geq0}x^{2n+1}/(2n+1)$, valid for
$|x|<1$.  The identity $2|M|p\leq M^2+p^2$ ensures $|b|\leq A_0$;
any zeros of the determinants require a limiting treatment of the logarithm.
Moreover, the pointwise series does not by itself justify exchanging a
small-$M$ expansion with the momentum integral near $p=0$.

To isolate the linear term without such an interchange, differentiate the
exact logarithm before setting $M=0$:
\begin{align}
\left.\frac{dV_{\rm odd}}{dM}\right|_{M=0}
 &=-\int\frac{d^2p}{(2\pi)^2}\,\frac{2\sin(p/\Lam)}{p}\nonumber\\
 &=-\frac1\pi\int_0^{\LUV}dp\,\sin(p/\Lam)\nonumber\\
 &=-\frac{\Lam}{\pi}
   \left[1-\cos(\LUV/\Lam)\right].
\label{eq:app-tadpole}
\end{align}
Here the radial integral employs the sharp $O(2)$-invariant momentum cutoff
$|p|\leq\LUV$.  Thus the linear term in $V_{\rm odd}$ is
$-M\Lam[1-\cos(\LUV/\Lam)]/\pi$.  In the mass-counterterm convention of
the main text, the required bare-mass magnitude is
\begin{equation}
 m_0=\frac{G\Lam}{\pi}\left[1-\cos(\LUV/\Lam)\right];
\end{equation}
its contribution to the potential must be assigned the opposite sign to
the tadpole.  The mass counterterm cancels the derivative at $M=0$, but
it does not cancel the higher odd dependence displayed in
Eq.~\eqref{eq:app-odd}.  Consequently, the renormalized unpaired
stationarity equation is not automatically identical to the paired one.

For a sharp radial cutoff, the coefficient of the linear term oscillates
as $\LUV/\Lam\to\infty$ and has no ordinary limit.  A different damping
prescription can yield a different regulated coefficient and must be
specified separately.  Quantitative statements concerning the residual
stationarity-equation difference require an explicit numerical comparison
of the two potentials with the same cutoff and counterterm convention.
Finally, for an even form factor $f(-\pslash)=f(\pslash)$, the two Dirac
determinants coincide, and this particular mass-odd discrepancy vanishes.

\subsection{Asymptotics and radial ultraviolet behavior}\label{app:asymp}
Subtract the local kernel from the Hermitian paired kernel:
\begin{align}\label{eq:app-subtraction}
\frac\pi G&=\int_0^{\Lambda_{\rm UV}}\frac{p\,dp}{p^2+M^2}
+\int_0^{\Lambda_{\rm UV}}p\,dp\,[K_f(p;M)-K_{\rm loc}(p;M)]\nonumber\\
&=\frac12\ln\left(1+\frac{\Lambda_{\rm UV}^2}{M^2}\right)
+\int_0^{\Lambda_{\rm UV}}p\,dp\,[K_f-K_{\rm loc}].
\end{align}
The second term is infrared finite as $M\to0$. Using
$K_f(p;0)=\cos(2p/\Lambda)/p^2$ and setting
$t=2p/\Lambda$, $s=\Lambda_{\rm UV}/\Lambda$, one obtains
\begin{equation}\label{eq:app-cin}
c_f(s)=\int_0^{2s}\frac{\cos t-1}{t}\,dt
=-\operatorname{Cin}(2s)
=\operatorname{Ci}(2s)-\gamma_E-\ln(2s).
\end{equation}
Consequently, for $M\ll\Lambda,\Lambda_{\rm UV}$,
\begin{equation}\label{eq:app-mass}
M_f\sim\Lambda_{\rm UV}\exp[c_f(s)-\pi/G].
\end{equation}
The asymptotic estimate obtained in Appendix~\ref{app:asymp} for the oscillatory case has
corrections of order $(M^2/\Lambda^2)\ln(\Lambda/M)$.
Since $\operatorname{Ci}(x)\to0$ at large $x$,
$M_f\sim\tfrac12e^{-\gamma_E}\Lambda e^{-\pi/G}$
for the specified radial-cutoff limit.
\begin{table}[b]
\caption{Cutoff diagnostic discussed in Appendix~\ref{app:asymp},
for $M=0.05\Lambda$. Only the local and Hermitian nonlocal kernels are compared.}
\label{tab:appUV}\begin{ruledtabular}
\begin{tabular}{ccc}
$\Lambda_{\rm UV}/\Lambda$ & local & Hermitian $f(\slashed{\partial})$\\
5&1.466&0.5479\\
20&1.907&0.5684\\
80&2.348&0.5628
\end{tabular}\end{ruledtabular}\end{table}
Table~\ref{tab:appUV} compares the cutoff dependence of the local kernel with the Hermitian nonlocal kernel used throughout this work.

\section{Numerical diagnostics and Lambert-$W$ roots}\label{app:diagnostics}
\subsection{Numerical checks}\label{app:num}
For the zero-temperature numerical diagnostics reported in this appendix, radial integrals were evaluated with adaptive
Gauss--Kronrod quadrature on logarithmically spaced intervals
between $M/10$ and $\Lambda_{\rm UV}$, resolving masses down to
$10^{-12}\Lambda$. Roots were bracketed and obtained using Brent's
method. An imposed infrared threshold $m_{\rm IR}$ at $M=0$
produces the cutoff-dependent $G_c$ values shown in
Table~\ref{tab:appIR}; these are not genuine critical couplings.
\begin{table}[t]\caption{Infrared-regulated values of $G_c$ with
$\Lambda_{\rm UV}=\Lambda$, as discussed in Appendix~\ref{app:num}.
The table includes only the local and Hermitian nonlocal models.}
\label{tab:appIR}\begin{ruledtabular}\begin{tabular}{ccc}
$m_{\rm IR}/\Lambda$& local & Hermitian $f(\slashed{\partial})$\\
$10^{-3}$&0.4548&0.5184\\
$10^{-6}$&0.2274&0.2423\\
$10^{-9}$&0.1516&0.1581
\end{tabular}\end{ruledtabular}\end{table}
\subsection{Lambert-$W$ roots and formal residues}\label{app:poles}
Under the analytic continuation used in the main text,
\begin{equation}\label{eq:app-factor}
\mathcal D_f(q)=
(q^2-M^2e^{2q/\Lambda})(q^2-M^2e^{-2q/\Lambda}).
\end{equation}
For the positive-energy branches,
\begin{align}
q=Me^{-q/\Lambda}&\Longrightarrow q=\Lambda W_0(M/\Lambda),\label{eq:app-w1}\\
q=Me^{q/\Lambda}&\Longrightarrow q=-\Lambda W_k(-M/\Lambda).
\label{eq:app-w2}
\end{align}
Here $W_0(x)=x-x^2+\frac32x^3+O(x^4)$, while
$W_0(-x)$ and $W_{-1}(-x)$ are real for $0<x<1/e$.
For $x\to0^+$,
\begin{equation}\label{eq:app-wh}
-W_{-1}(-x)=\ln(1/x)+\ln\ln(1/x)+\cdots.
\end{equation}
At a formal pole of $h(q)=q-Me^{q/\Lambda}$,
\begin{equation}\label{eq:app-res}
h'(q_0)=1-q_0/\Lambda,
\qquad \operatorname{Res}_{q=q_0}h(q)^{-1}
=\frac1{1-q_0/\Lambda}.
\end{equation}
Thus the heavier $W_{-1}$ root has the opposite scalar
residue sign; this does not independently establish a physical ghost.
The numerical roots associated with Eqs.~\eqref{eq:app-w1} and \eqref{eq:app-w2} for
$M=0.1\Lambda$ are
$(q_-,q_+,q_h)/\Lambda=(0.091277,0.111833,3.577152)$;
the same formal analytic continuation also yields complex roots, including
$(4.449-7.307i)\Lambda$ for $k=1$.
The analysis of formal roots is restricted to the Hermitian exponential form factor defined in the main text.
\endgroup


\begin{thebibliography}{99}
\bibitem{BasBeneito2023} A. Bas i Beneito, G. Calcagni, and L. Rachwa\l, in \emph{Handbook of
Quantum Gravity}, edited by C. Bambi, L. Modesto, and I. Shapiro (Springer, Singapore, 2023),
arXiv:2211.05606.
\bibitem{Yukawa} H. Yukawa, Proc. Phys. Math. Soc. Jpn. \textbf{17}, 48 (1935);
Phys. Rev. \textbf{77}, 219 (1950).
\bibitem{Efimov1967} G. V. Efimov, Commun. Math. Phys. \textbf{5}, 42 (1967).
\bibitem{Brekke1988} L. Brekke, P. G. O. Freund, M. Olson, and E. Witten, Nucl. Phys.
\textbf{B302}, 365 (1988).
\bibitem{Frampton1988} P. H. Frampton and Y. Okada, Phys. Rev. Lett. \textbf{60}, 484 (1988).
\bibitem{Eliezer1989} D. A. Eliezer and R. P. Woodard, Nucl. Phys. \textbf{B325}, 389 (1989).
\bibitem{Ghoshal2000} D. Ghoshal and A. Sen, Nucl. Phys. \textbf{B584}, 300 (2000).
\bibitem{Evens1991} D. Evens, J. W. Moffat, G. Kleppe, and R. P. Woodard, Phys. Rev. D
\textbf{43}, 499 (1991).
\bibitem{Tomboulis1997} E. T. Tomboulis, arXiv:hep-th/9702146.
\bibitem{Tomboulis2015} E. T. Tomboulis, Phys. Rev. D \textbf{92}, 125037 (2015).
\bibitem{Buoninfante2019} L. Buoninfante, G. Lambiase, and A. Mazumdar, Nucl. Phys.
\textbf{B944}, 114646 (2019).
\bibitem{Barnaby2008} N. Barnaby and N. Kamran, J. High Energy Phys. 02 (2008) 008.
\bibitem{Calcagni2008} G. Calcagni, G. Nardelli, and M. Montobbio, Phys. Lett. B \textbf{662},
285 (2008).
\bibitem{Modesto2021} L. Modesto and G. Calcagni, J. High Energy Phys. 10 (2021) 169.
\bibitem{Nascimento2026} J. R. Nascimento, G. J. Olmo, A. Yu. Petrov, P. J. Porfirio, and
R. N. da Silva, Phys. Rev. D \textbf{113}, 045015 (2026), arXiv:2505.07483.
\bibitem{Krasnikov}
N.~V.~Krasnikov,
Theor. Math. Phys. \textbf{73}, 1184-1190 (1987).
\bibitem{Gabriel}
G.~Q.~Garcia, P.~J.~Porf{\'\i}rio, E.~Brito and D.~C.~Moreira,
[arXiv:2610.27349 [hep-ph]].
\bibitem{Porfirio:2026zga}
P.~J.~Porf{\'\i}rio and R.~N.~da Silva,
[arXiv:2610.06390 [hep-th]].
\bibitem{Moffat}
J.~W.~Moffat,
Phys. Rev. D \textbf{41}, 1177-1184 (1990).
\bibitem{Thirring1958} W. E. Thirring, Ann. Phys. (N.Y.) \textbf{3}, 91 (1958).
\bibitem{NJ} Y. Nambu and G. Jona-Lasinio, Phys. Rev. \textbf{122}, 345 (1961);
\textbf{124}, 246 (1961).
\bibitem{Klevansky} S. P. Klevansky, Rev. Mod. Phys. \textbf{64}, 649 (1992).
\bibitem{Belchior}
F.~M.~Belchior, J.~R.~Nascimento, A.~Y.~Petrov and P.~J.~Porfirio,
[arXiv:2607.28160 [hep-th]].
\bibitem{GrossNeveu1974} D. J. Gross and A. Neveu, Phys. Rev. D \textbf{10}, 3235 (1974).
\bibitem{MosheZinnJustin2003} M. Moshe and J. Zinn-Justin, Phys. Rep. \textbf{385}, 69 (2003).
\bibitem{Coleman1973} S. Coleman, Commun. Math. Phys. \textbf{31}, 259 (1973).
\bibitem{Coleman1975} S. Coleman, Phys. Rev. D \textbf{11}, 2088 (1975).
\bibitem{DashenFrishman1975} R. Dashen and Y. Frishman, Phys. Rev. D \textbf{11}, 2781 (1975).
\bibitem{Witten1978} E. Witten, Nucl. Phys. \textbf{B145}, 110 (1978).
\bibitem{Wilson} K. G. Wilson, Phys. Rev. D \textbf{10}, 2445 (1974).
\bibitem{Stra}
R. L. Stratonovich,
Sov. Phys. Dokl. \textbf{2}, 416 (1958).
\bibitem{Hubbard}
J. Hubbard,
Phys. Rev. Lett. \textbf{3}, 77 (1959).
\end{thebibliography}
\end{document}